# Forecasting Hyponatremia in hospitalized patients Using Multilayer Perceptron and Multivariate Linear Regression Techniques

Prasannavenkatesan Theerthagiri

Department of Computer Science and Engineering, School of Technology, GITAM University, Bengaluru-561203, India, email: vprasann@gitam.edu

**Abstract:** The percentage of patients hospitalized due to hyponatremia is getting higher. Hyponatremia is the deficiency of sodium electrolyte in the human serum. This deficiency might indulge adverse effects and also associated with longer hospital stay or mortality, if it wasn't actively treated and managed. This work predicts the futuristic sodium levels of patients based on their history of health problems using multilayer perceptron (MLP) and multivariate linear regression (MLR) algorithm. This work analyses the patient's age, information about other disease such as diabetes, pneumonia, liver-disease, malignancy, pulmonary, sepsis, SIADH, and sodium level of the patient during admission to the hospital. The results of the proposed MLP algorithm is compared with MLR algorithm based results. The MLP prediction results generates 23-72 % of higher prediction results than MLR algorithm. Thus, proposed MLR algorithm has produced 57.1 % of reduced mean squared error rate than the MLR results on predicting future sodium ranges of patients. Further, proposed MLR algorithm produces 27-50 % of higher prediction precision rate.

## 1. Introduction

Once the human population is suffered by life-threatening diseases such as stroke, central nervous diseases, chronic diseases, renal electrolytes imbalance, then the possibility of occurrence of the same or indulging side effects is higher. The poor diagnosis, treatments or other health problems of the patients might increases the chances of occurrence of the same for the second time. These kinds of deadly diseases should be actively overserved and managed with intensive care.

The homeostatic levels of the human body should be maintained within the normal range [1]. The healthy adult human body should have the following range; the serum sodium (Na+) as 136-145 mmol/L; serum potassium (K+) as 3.3-4.5 mmol/L; serum chloride (Cl-) as

96-108 mmol/L; urine sodium as 40-220 mmol/day; urine potassium as 25-125 mmol/day; urine serum chloride as 110-250 mmol/day [2, 3]. Any abnormalities in these recommended range might indulge adverse effects in the human body and its severity is based on the level of homeostatic concentration or dilution.

The inappropriate antidiuretic hormone secretion might leads to cause the rapid loss of sodium in the serum of human. The loss of serum sodium concentration in human blood is called as hyponatremia. The sodium level in the blood is diluted by the excess water intake and frequent urination. Typically, the hyponatremia is a common disorder and is being observed in many hospitalized patients [4]. It can lead to mortality of the patient, if the required essential treatments is not given. Even with the adequate intensive treatments the percentage of mortality is high for the patients with hyponatremia [5]. The hyponatremia also found in patients with the excessive Arginine Vasopressin (AVP) hormone in the plasma of the blood.

The hyponatremia can be categorized into acute and chronic based on the concentration rate of the serum sodium. The too lower serum sodium concentration is very deleterious and it can affect the normal functions of other organs of the human body [6]. Typically, the most frequent hyponatremia might leads to serious complications in the central nervous system. The patients with acute hyponatremia should be given with intensive and prompt treatments; otherwise, it might lead to life-threatening. The chronic hyponatremia develops the non-renal diseases in other organs of the human body, which often increases the morbidity and mortality.

The poor prognosis and treatments given to the patients with hyponatremia leads to increase the length of hospital stay, especially for the heart patients [7]. The hyponatremia might be managed optimally with prompt treatments to the patients, which may increase the serum sodium concentration and reduces the severity and in-hospital mortality. The prompt diagnosis and optimal recognized treatments of hyponatremia to the patients might be helpful for improving their body conditions to normal and also allows them to the normal hospitalization, reduces the duration of hospitalization and hospitalization costs as well [8].

Increasing the serum sodium concentration too rapidly might leads to life-threatening diseases and sequel side effects. The serum sodium must be increased by recommended level only; otherwise it increases the possibility of fatal heart or neurologic disorders [9]. One of the major therapies provided to the hyponatremia is associated with the blocking the actions of the arginine vasopressin receptors to resist from elimination of electrolytes during the urination; it might raise up the serum sodium concentration [10].

The serum sodium concentration should be managed within the recommended range; also based on the homeostatic mechanisms, AVP, renal water excretion, and thirst. Any abnormal changes with the electrolytes of blood, renal chemical levels, and water level could lead to hyponatremia. Even, a small unbalance in the renal functions, which was not treated carefully could increase the possibility of consistent morbidity or on severe condition may be death as well [11, 12]. The adverse effects of the hyponatremia on the human population motivates the research work on prediction of future sodium ranges and possibility of readmission due to hyponatremia on patients. This research work uses the regression and neural networks for the analysis of sodium ranges.

The remainder of the section is organized as follows. The section 2 reports the existing literature works about hyponatremia. The symptoms, causes, existing diagnosis, treatments, effects of hyponatremia. The section 3 explains the proposed methodology and its stepwise details to predict the future sodium ranges for the patients under the hyponatremia treatment. The prediction results analysis and evaluation of the proposed methodology is presented in the section 4. Finally, section 5 concludes with future enhancements.

**2.     Literature survey**

This section summarizes the brief review of existing literacy works that addresses the importance of hyponatremia and appropriate treatments. Numerically the hyponatremia is defined as the incident that the serum sodium concentration (Na +) reduces below to 136 (normal range 136–145) mEq/l. This type of disorder is common in many hospitalized patients. As mentioned earlier, the excessive water retention in the body or very frequent urination causes the disorder of dilutional hyponatremia [13]. The water intake should not exceed the kidneys' excretory capacity. If exceeds, it causes the sodium dilution in the serum, which may lead to hyponatremia, hypo-tonicity, and hypo-osmolality [14].

Typically, the patients will drink more water, if they are affected by syndrome of inappropriate antidiuretic hormone secretion (SIADH); since the SIADH patients always feels thirst. The treatments and medications are provided to the patients with hyponatremia based on their age, severity, hormone conditions, renal dysfunction, chronic disease, adrenal level, and nervous system [15, 16]. A variety of drugs associated with the treatment of hyponatremia are sulfur-containing diuretics drugs, vasopressin receptor antagonist drugs, tolvaptan, desmopressin, etc.; these drugs might reduce the risks related to the hyponatremia. The falsely prognosis of low serum sodium concentration is termed as pseudohyponatremia. It can happens, when the patients having extreme hyperlipidemia or hyperproteinemia and it is identified by flame photometry or indirect potentiometry [17].

In most of the cases, the excessive AVP secretion with the absence of elevated plasma osmolality is the major issue that causes the hyponatremia. The kidney may retain the water because of the elevated AVP secretion. Thus, the reduced water excretion and increased AVP concentration have direct association with them [18]. The patients having too-much AVP secretion and exceeding their water intake by 800 ml/day might cause the water retention; and also it dilutes the fluid compartments, which leads to cause hyponatremia.

Based on the age, the occurrence of hyponatremia increases and also the documented studies suggests that the old aged people have higher percentage; such that 60-aged population have 53% of this disorder in a year. Further, SIADH disorder is most common in majority of hospitalized patients with hyponatremia [19]. The valproic acid plays a major role on controlling the Na+ channels. It is an 8-carbon 2-chain fatty acid and human's liver metabolizes it [20]. The valproic acid mediates for the management and recovery of serum sodium concentration through the depolarization of spinal cord and cortical neurons for the patients.

Typically, its therapeutic range should not exceed 50-100 mcg/mL per dosage [21]. While its range produce positive responses with minimal side effects experienced by the patients. Also, the level of valproic acid in the blood level should be monitored and altered within the recommended range based on the patients' responses in a less frequent but regular interval. The time taken to make effects by the valproic acid on the human body varies between the patients [22].

The subtherapeutic/supratherapeutic levels of valproic acid to patients, might make them to the risk conditions or indulges toxic side effects. If required, the additional testing/treatments should be given to the patients based on medication's effectiveness/ineffectiveness, patient's side effects, effects on the central nervous system, and other complications faced by the patients [23]. Further, the patients with multiple medications should be taken additional care with continuous evaluation of valproic acid level because of the interaction of other medications. Also, protein level in the blood also essentially be monitored because supratherapeutic valproic acid level has more influence on it [24].

The research gap in predicting future sodium range of the patients motivates this research work and analysis of possibility of readmission due to hyponatremia.

**3.    Methodology**

This section describes the methodologies for the prediction of the future sodium ranges of the patients affected by the hyponatremia. The multilayer perceptron and multivariate linear regression techniques are adopted in this work for the prediction of future values of the sodium.

The proposed methodology makes the analyses and predicts the futuristic sodium levels of patients based on their history of health conditions. This research work also facilitates the way to forecast the possibility of the occurrence of hyponatremia once again using artificial intelligence-based algorithms. Further, to emulate the futuristic healthcare of the patients based on their previous history of illness/diseases, the proposed research work had developed a MLR and MLP based future health prediction algorithm.

### 3.1 Dataset

The dataset this research is obtained from Cerner Health Facts database. This is a dataset of hospitalized patients during January 2000 and November 2014 collected from various clinics and hospitals in the United States [9]. The dataset contains 1048576 number of patients' information. The dataset contains features of patients such as patient's length of stay, hypertension, coronary_artery_disease, heart_failure, chronic_kidney_disease, end_stage_renal_disease, cirrhosis, chronic_liver_disease, copd, lung_cancer, adrenal_insufficiency, hypothyroidism, depression, dementia, myocardial infarction, peripheral vascular disease, cerebrovascular disease, rheumatologic disease, peptic ulcer disease, metastatic cancer diabetes, diabetes_complication, hemiplegia, glucose level during 24 hr before admission, glucose levels during 24 hr before admission excluding values <=20 and >2000, sodium levels during 24 hr before admission, race, age, gender, sodium levels adjusted for glucose, serum sodium categories, pneumonia , malignancy, pulmonary, sepsis, SIADH, and outcomes [9].

The details of the patients such as age ($A$), gender ($G$), information about diabetes ($D$), pneumonia ($P$), liver-disease ($L$), malignancy ($M$), pulmonary ($Pu$), sepsis ($Se$), SIADH ($S$), and sodium level ($Na$) during admission are taken for this research analysis. The patients are grouped into four categories based on their ages. The age groups are (i) 18 to <45, (ii) 45 to <65, (iii) 65 to <75, (iv) ≥75 years old [9]; the sodium range is categorized as 1 for < 120; 2 for ≥ 120 to < 125; 3 for ≥ 125 to < 130; 4 for ≥ 130 to < 135; 5 for ≥ 135 to < 138; and 6 for ≥ 138 to < 140. Similarly, based on gender the patients are grouped namely male and female for the training and learning by the MLR and MLP algorithms.

The data preprocessing and cleaning process removes the missing and outliers data values from the dataset. The resulted dataset after preprocessing is reduced to one million patient records with the above listed 10 required relevant features of patient details. In this dataset, approximately about 49658 records are missing the essential required patient details. The numerical features from the dataset are taken as the input attributes (gender, age, diabetes, pneumonia, liver_disease, malignancy, pulmonary, sepsis, SIADH) and one feature (Sodium range) is considered as the output attribute. A sample of patient's information is presented in Table.1.

**Table.1 Sample record of cleaned dataset**

| Gender | Age | Diabetes | Pneumonia | Liver Disease | Malignancy | Pulmonary | Sepsis | Sodium Range |
|---|---|---|---|---|---|---|---|---|
| 1 | 1 | 1 | 0 | 0 | 1 | 0 | 0 | 1 |
| 1 | 1 | 0 | 0 | 0 | 0 | 1 | 0 | 3 |
| 1 | 2 | 0 | 0 | 0 | 0 | 0 | 0 | 3 |
| 1 | 0 | 1 | 0 | 0 | 0 | 0 | 0 | 3 |
| 0 | 1 | 0 | 0 | 0 | 0 | 0 | 0 | 6 |
| 1 | 2 | 0 | 0 | 0 | 0 | 0 | 0 | 3 |
| 1 | 1 | 0 | 0 | 0 | 1 | 0 | 0 | 4 |
| 0 | 0 | 0 | 0 | 0 | 0 | 0 | 0 | 3 |
| 1 | 3 | 0 | 0 | 0 | 0 | 0 | 0 | 3 |
| 1 | 2 | 0 | 0 | 0 | 0 | 0 | 0 | 3 |

### 3.2 Multivariate Linear Regression (MLR) Algorithm

The multivariate linear regression is the knowledge based learning algorithm, works based on training of the dataset. The MLR can be defined as improved knowledge driven expert system [25]; it generates linear hypothesis and determines the weights for the given variable based on the learning process and parameters [26]. The multivariate linear regression algorithm is a statistics based model. The MLR algorithm describes the relationship between two or more dependent and independent variables based on the given dataset [27].

The MLR algorithm computes and analyses the regression for producing the optimized or appropriate results [28]. It also determines the correlation and assesses the testing/validation and usefulness of the model using various stages of multivariate linear regression model [28, 29]. The MLR can be defined as denoted in the equation (1).

$$MLR_i(a,b,k) = \varphi + \alpha_1 A + \alpha_2 G + \alpha_3 D + \alpha_4 P + \alpha_5 L + \alpha_6 M + \alpha_7 Pu + \alpha_8 Se + \alpha_9 S + \alpha_{10} Na \qquad (1)$$

In the equation 1, $\alpha_1, \ldots\ldots\ldots, + \alpha_{10}$ are the set of additive predictor functions, $\varphi$ is the intercept associated with each functions [30], the variable *'A'* represents the patient's age, The gender of the patient is denoted as *'G'*, the variable *'D'* is denotes the patients with diabetes, patients with pneumonia is given as *'P'*, *'L'* gives the patients with liver-disease, *'M'* is malignancy, *'Pu'* is pulmonary, *'Se'* is sepsis, *'S'* is SIADH, *'Na'* is the sodium level of the patients during admission to the hospital.

### 3.3    Multilayer Perceptron (MLP)

This work proposes and develops the MLP based future sodium prediction algorithm. The MLP algorithm works using the artificial neural network. Typically, in a simple three-layer network, there will be input layer in the first, hidden layer in the middle, and output layer in the last. In the input layers, the input dataset will be feed, while the output layer contains the output, which is generated based on the dataset given to the input layer and computations from the hidden layer. The number of hidden layer can be varied (increased/decreased) based on the complexity of the given problem [37]. In general the main objective of any neural network model is to optimize or approximate the given function *f(x)*. Similarly, the multilayer perceptron neural network model will find the best optimized or approximate solution to the given complex problem by using techniques such as classification, regression, or mapping functions [31].

In the proposed MLP based supervised learning technique, each neuron uses a nonlinear activation function and backpropagation for training. The proposed MLP network consists of several chained functions. Let consider, a classifier problem *y= f(x)*; here the output *y* is driven by the input *x* and its corresponding mapping solution given by MLP based on the best approximation of the given classifier function. Such that the MLP computes the best optimized solution as *y = f(x; θ)*; where *θ* is the learning parameter of given problem. For example, the three layer MLP network can be formulated as *f(x) = f (x3) (f (x2) (f (x1) (x)))*. In each layer, the MLP performs several defined transformation and/or linear summation functions with the inputs. In MLP, each of these layers are symbolized as *y = f (W.x.T + b)*; where the activation functions is denoted as *f*, the weights or set of parameter of the problem are indicated as *w*, the variable *x* is the input, and *b* represents the bias vector [32].

In the proposed MLP algorithm, the output of previous layer is the input to the next layer. Such that the layers of the MLP of fully connected with each other layers in the network.

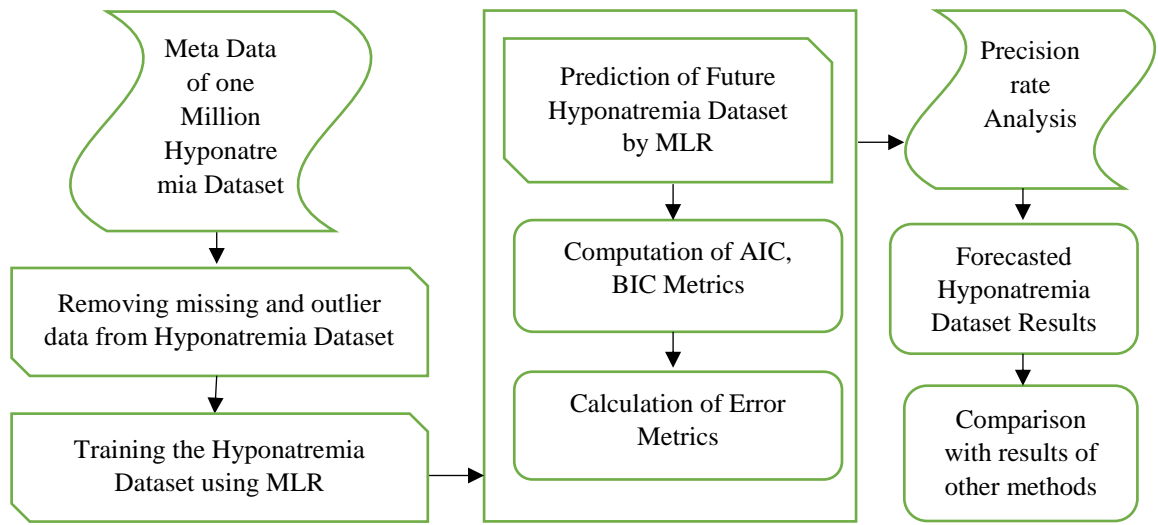

**Fig. 1. Block diagram of the proposed model**

Thus, each unit functions of the layer are always connected to the all other layers' unit function in the neural network. Although, each layer's unit functions (*i.e.* weights and other set of parameters) are independent to other layer's unit functions. It means the weights of the each layer's unit functions are unique. Further, the MLP network defines the loss function, which can measure the performance (sodium prediction) of the proposed MLP's classification technique. When the loss function has high value, the MLP doesn't made accurate classification or prediction solution to the given problem and otherwise it is vice versa [31, 32].

Fig.1 depicts the flow of the proposed MLP based future sodium prediction algorithm. Firstly, one million patients hyponatremia dataset collected from the hospitals. The outlier and missing data are removed from the hyponatremia dataset using imputation process. Then the 0.5 million of hyponatremia dataset is given to the input layer as the input and it is trained by using the multilayer perceptron algorithm. To obtain the better optimal prediction model, the number hidden layers are varied from two to twenty by the unit of two. This MLP learning and training process gives the prediction results in the output layer.

The prediction results are evaluated for the quality check of the prediction by using the Akaike Information criterion (AIC) and Bayesian Information criterion (BIC). Again, the error performance metrics such as Mean Squared Error (MSE), Root Mean Squared Error (RMSE), Mean Absolute Error (MAE), Mean Absolute Relative Error (MARE), and Root Mean Squared Relative Error RMSRE (RMSRE) are computed for the prediction results. The AIC, BIC analysis and error performance metrics computation results us the best quality future sodium prediction dataset [33, 38]. The performance results of the future sodium prediction dataset are analyzed with the precision rate and compared with other existing results.

## 4. Results Evaluation and Discussion

This section gives the detailed results analysis, evaluation of performance metrics, and comparative result analysis. The taken dataset has been split into 70 %, 15%, and 15% for the training, validation, and testing respectively. To evaluate and validate the performance of the machine learning model resampling methods are adopted. This method estimates the prediction ability on the machine learning algorithm on new unseen input data. In this work, the 'k' value is chosen as 10, therefore, it can be called as a 10-fold cross-validation resampling method. The 10-fold cross-validation method intends to reduce the bias of the prediction model.

### 4.1 Results of MLR Algorithm

In order to predict the future sodium range, the multivariate linear regression algorithm trains the 0.5 million dataset containing the patients' information such as age (*A*), gender (*G*), information about diabetes (*D*), pneumonia (*P*), liver-disease (*L*), malignancy (*M*), pulmonary (*Pu*), sepsis (*Se*), SIADH (*S*), and sodium level of the patients during admission to the hospital (*Na*). The MLR algorithm trains the 0.5 million patients dataset and performs the testing and validation operations on the given dataset and finally produces the dataset for the future sodium range values.

In order to determine the quality dataset for the prediction of the future sodium values, generalized regression parameters such as *(a, b, k)* are varied; where the parameter *a* is the number of input channels, *b* is the number of output channels and *k* represents the delay value. The *(a, b, k)* parameters are varied as (1,2,1), (1,2,2), (2,2,1), (2,2,2), (2,2,3), (2,3,2), (2,3,3), (3,2,1), (3,2,2), and (3,3,2). Table 2, summarizes the results of error performance metrics such as MSE, RMSE, MAE, MARE, and MSRE for the MLR parameters *(a, b, k)* using the MLR algorithm [33].

As per definition of MSE, the resultant lowest MSE value among the different *(a, b, k)* parameter gives the feasible realistic solution. Therefore in Table 2, the lowest MSE value that is *(a, b, k)* parameter *(2,2,1)* are highlighted with boldface. The error performance metric values for the MSE, RMSE, MAE, MARE, and MSRE for the MLR parameters *(2,2,1)* are 0.1894, 0.4352, 0.4162, 0.1036, and 0.1121 respectively; it is the lowest among other *(a, b, k)* parameters. Therefore, the corresponding dataset of the *(2,2,1)* parameter is considered as the appropriate and optimistic solution for the given hyponatremia patient dataset.

**Table.2 Performance Error Metrics resulted by MLR**

| Error Metrics/ (a, b, k) | MSE | RMSE | MAE | MARE | RMSRE |
|---|---|---|---|---|---|
| (1,2,1) | 0.215 | 0.4637 | 0.4386 | 0.107 | 0.1148 |
| (1,2,2) | 0.2643 | 0.5141 | 0.4849 | 0.1173 | 0.1247 |
| **(2,2,1)** | **0.1894** | **0.4352** | **0.4162** | **0.1036** | **0.1121** |
| (2,2,2) | 0.2668 | 0.5165 | 0.4908 | 0.1192 | 0.1263 |
| (2,2,3) | 0.2923 | 0.5406 | 0.5097 | 0.123 | 0.1303 |
| (2,3,2) | 0.2585 | 0.5084 | 0.4794 | 0.1161 | 0.1236 |
| (2,3,3) | 0.5028 | 0.7091 | 0.6688 | 0.1611 | 0.1699 |
| (3,2,1) | 0.2152 | 0.4639 | 0.439 | 0.1071 | 0.1149 |
| (3,2,2) | 0.2671 | 0.5168 | 0.491 | 0.1193 | 0.1264 |
| (3,3,2) | 0.2585 | 0.5084 | 0.4795 | 0.1161 | 0.1236 |

In addition, the results of AIC and BIC metrics for the MLR algorithm given in Table 3 confirms that the *(2,2,1)* parameter produces the lowest metric results. As per the AIC and BIC definition, the better quality and stable can be given by the lowest AIC or BIC valued corresponding dataset. In Table 3, the boldfaced *(2,2,1)* parameter gives the lowest AIC and BIC metrics such as 2211.489 and 2214.856 respectively. The resultant values of AIC, BIC, and error metrics confirms that the dataset of *(2,2,1)* parameter produces better prediction results. Therefore, the parameter *(2,2,1)*'s corresponding dataset is taken as the result of the future sodium prediction values.

**Table.3 Results of AIC and BIC Metrics for MLR**

| Criterion/ (a, b, k) | AIC | BIC |
|---|---|---|
| (1,2,1) | 731809.6 | 731776.2 |
| (1,2,2) | 811329.74 | 813626.37 |
| (2,2,1) | 591230.55 | 593907.18 |
| (2,2,2) | 611590.781 | 662324.148 |
| (2,2,3) | 672218 | 672351.3 |
| (2,3,2) | **600211** | **600215** |
| (2,3,3) | 768794.9 | 768828.3 |
| (3,2,1) | 733165.4 | 733132.1 |
| (3,2,2) | 720541.85 | 780735.22 |
| (3,3,2) | 822211.59 | 901378.22 |

## 4.2 Results of MLP Algorithm

The multilayer perceptron algorithm trains the dataset containing the 0.5 million patients' information such as age (*A*), gender (*G*), information about diabetes (*D*), pneumonia (*P*), liver-disease (*L*), malignancy (*M*), pulmonary (*Pu*), sepsis (*Se*), SIADH (*S*), and sodium level (*Na*) of the patients during admission to the hospital. In order to determine the quality dataset for the prediction of the future sodium values, the number of hidden neurons are varied from two to twenty [34, 36]. Table. 4 summarizes the resultant performance error metrics values such as MSE, RMSE, MAE, MARE, and MSRE for the MLR algorithm with several different neurons.

**Table.4 Performance Error Metrics resulted by MLP**

| Metrics/Neurons | MSE | RMSE | MAE | MARE | RMSRE |
|---|---|---|---|---|---|
| 2 | 0.052 | 0.2281 | 0.1521 | 0.0418 | 0.073 |
| 4 | 0.1012 | 0.3181 | 0.2791 | 0.071 | 0.0919 |
| **6** | **0.0227** | **0.1505** | **0.0691** | **0.0196** | **0.0441** |
| 8 | 0.0942 | 0.3069 | 0.273 | 0.07 | 0.0898 |
| 10 | 0.0563 | 0.2372 | 0.1693 | 0.0492 | 0.1017 |
| 12 | 0.1092 | 0.3304 | 0.2975 | 0.0752 | 0.0956 |
| 14 | 0.1014 | 0.3184 | 0.2832 | 0.0728 | 0.0967 |
| 16 | 0.1018 | 0.3191 | 0.2862 | 0.0727 | 0.0914 |
| 18 | 0.0806 | 0.2839 | 0.248 | 0.0636 | 0.0773 |
| 20 | 0.1005 | 0.317 | 0.2848 | 0.0722 | 0.0894 |

In Table. 4, the lowest MSE value is highlighted as neuron 6. The error performance metric values for the MSE, RMSE, MAE, MARE, and MSRE for the neuron 6 by MLP algorithm are 0.0227, 0.1505, 0.0691, 0.0196, and 0.0441 respectively; it is the lowest among other neurons. Therefore, the corresponding dataset of the neuron 6 is considered as the appropriate and optimistic solution for the given hyponatremia patient dataset.

Moreover, the results of AIC and BIC metrics for the MLP algorithm is summarized in Table 5. It confirms that the neuron 6 generates the lowest metric results. In Table 3, the boldfaced (2,2,1) parameter gives the lowest AIC and BIC metrics such as 525639.24 and 525605.88 respectively. The resultant values of AIC, BIC, and error metrics confirms that the dataset of neuron 6 produces better prediction results. Therefore, the neuron 6's corresponding dataset is considered as the result of the future sodium prediction values in this scenario.

**Table.5 Results of AIC and BIC Metrics for MLP**

| Criterion/Neurons | AIC | BIC |
|---|---|---|
| 2 | 950044.093 | 950021.848 |
| 4 | 759348.725 | 759326.48 |
| **6** | **525639.24** | **525605.88** |
| 8 | 780974.062 | 780951.818 |
| 10 | 950819.348 | 950797.103 |
| 12 | 708582.641 | 708560.397 |
| 14 | 777258.361 | 777236.116 |
| 16 | 764183.817 | 764161.572 |
| 18 | 832843.723 | 832821.478 |
| 20 | 757474.068 | 757451.823 |

### 4.3 Result analysis for MLR and MLP algorithm

The future sodium prediction results obtained by the techniques MLR and MLP are compared for the analysis of accurate prediction of results. Fig. 2 gives the comparative future sodium prediction results by using the techniques MLP and MLR for the critical hyponatremia patients. The observed and predicted results of patient's serum sodium range such as less than 120, 120 to 125, and 126 to 130 are depicted in the Fig.2. For the sodium range less than 120 category, the total number of observed patients are 2568; the proposed MLP algorithm had predicted the total number of patients under less than 120 category as 2537; whereas, the MLR had predicted it as 5399 patients for the less than 120 category.

The proposed MLP algorithm had produced higher accuracy of prediction, in which the prediction difference with the observed results is 1.21 % only; whereas the MLR algorithm has prediction difference with the observed results as 71 %. The total number of observed patients for the sodium range 120 to 125 category is 9639; the proposed MLP algorithm had predicted the total number of patients for the 120 to 125 category as 10554; whereas, the MLR had predicted it as 15370 patients for the same category. In this case, the prediction difference with the observed results MLP and MLR algorithms are 9 and 45 % respectively.

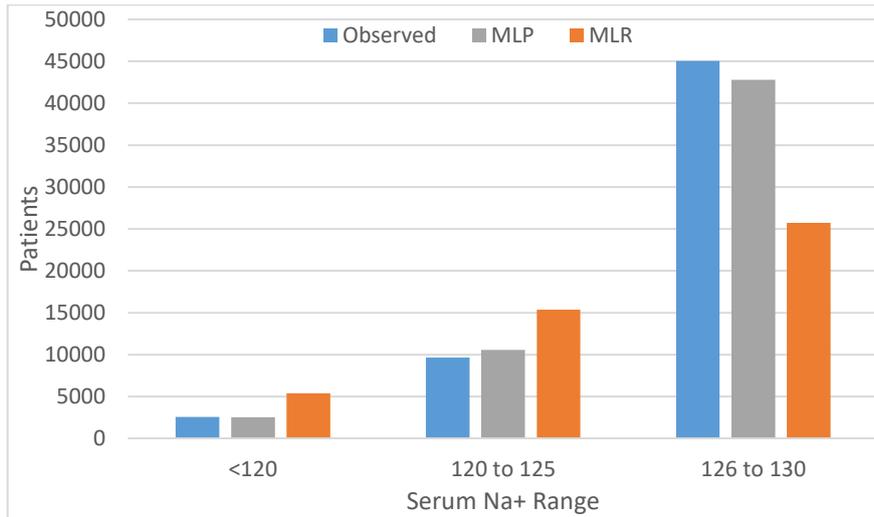

**Fig. 2. Comparison of Na+ prediction results by MLP and MLR (Critical)**

Similarly, For the sodium range 126 to 130 category, the total number of observed patients are 45024; the proposed MLP algorithm had predicted the total number of patients for the 126 to 130 category as 42797; whereas, the MLR had predicted it as 25711 patients for the same category. Correspondingly, the MLP algorithm has lower prediction difference with the observed results as 5 %; whereas, the MLR algorithm has higher prediction difference with the observed results as 54.6 %.

For the stable hyponatremia patients, the comparative future sodium prediction results by using the techniques MLP and MLR are depicted in Fig. 3. The total number of observed patients for the sodium range 131 to 135 category is 300187; the proposed MLP algorithm had predicted the total number of patients for the same category as 305875; whereas, the MLR had predicted it as 385388 patients for this category. In this case, the prediction difference with the observed results MLP and MLR algorithms are 1.8 and 24.8 % respectively.

Similarly, For the sodium range 136 to 138 category, the total number of observed patients are 142582; the proposed MLP algorithm had predicted the total number of patients for the same sodium category as 13827; whereas, the MLR had predicted it as 68117 patients for the same category. In the same way, the MLP algorithm has lower prediction difference with the observed results as 3.1 %; whereas, the MLR algorithm has higher prediction difference with the observed results as 70.6 %.

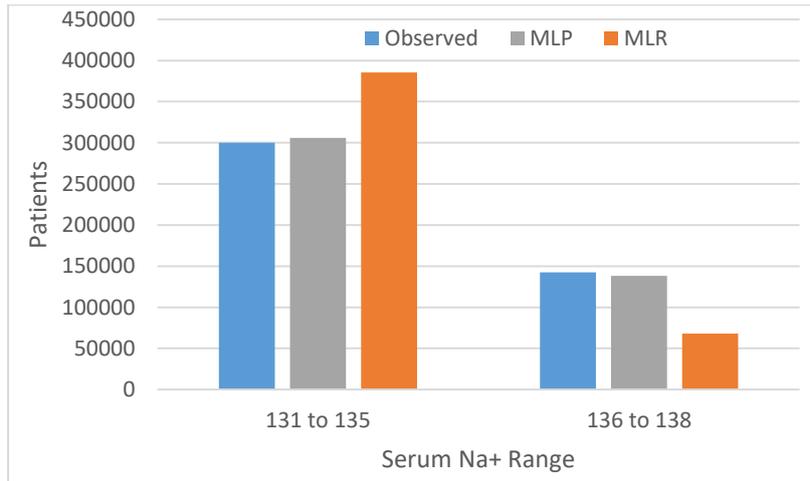

**Fig. 3. Comparison of Na+ prediction results by MLP and MLR (Stable)**

Table.6 gives the best results of performance error metric for the proposed MLP and MLR algorithms. The appropriate stable dataset *(2,2,1)* discovered by the MLR algorithm has the performance error metrics such as MSE, RMSE, MAE, MARE, and MSRE as 0.1894, 0.4352, 0.4162, 0.1036, and 0.1121 respectively. Similarly, quality dataset *(neuron 6)* determined by the proposed MLP algorithm has the performance error metrics such as MSE, RMSE, MAE, MARE, and MSRE as 0.0227, 0.1505, 0.0691, 0.0196, and 0.0441 respectively. Henceforth, these confirms that the proposed MLP algorithm had generated the lower error rates than the MLR error rates.

**Table.6 Error Metrics result analysis for MLR and MLP**

| Metrics/Algorithm | MSE | RMSE | MAE | MARE | RMSRE |
|---|---|---|---|---|---|
| **MLR** | 0.1894 | 0.4352 | 0.4162 | 0.1036 | 0.1121 |
| **MLP** | 0.0227 | 0.1505 | 0.0691 | 0.0196 | 0.0441 |

Fig.4. gives the patients' age wise hyponatremia prediction results based on the MLP algorithm. From this pie chart, it can be seen that the patients with below the age of 45 are getting affected by the hyponatremia is very less (i.e.) 17 %. When the patients' age is above

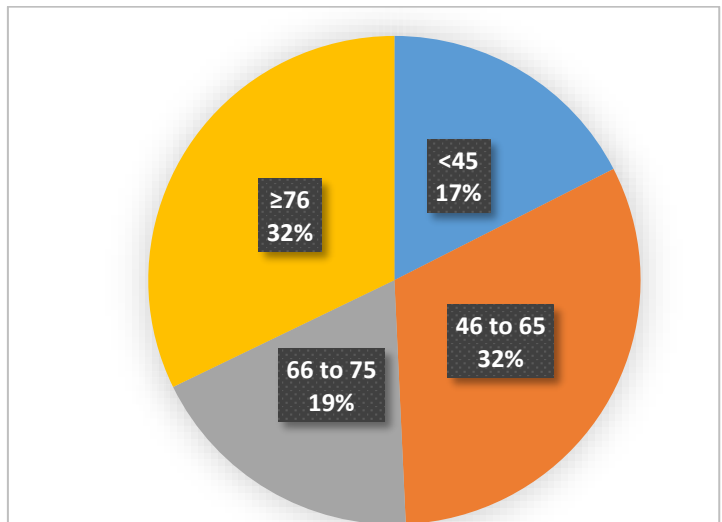

**Fig. 4. Age wise hyponatremia prediction by MLP**

75 or the patients' age is 46 to 65 then the number of patients getting affected by the hyponatremia is very high (i.e.) 32 %. In between the age of 66 to 75 there are 19 % of patients had treatment for hyponatremia. Fig.5 illustrates the pie chart analysis on MLP based hyponatremia prediction results based on gender. From the Fig. 5, it is clear that female patients are highly suffered by hyponatremia than the male patients. Such that 54 % of female patients had treatments for the hyponatremia, whereas 46 % of male patients only had treatments for hyponatremia from the dataset of 0.5 million patients.

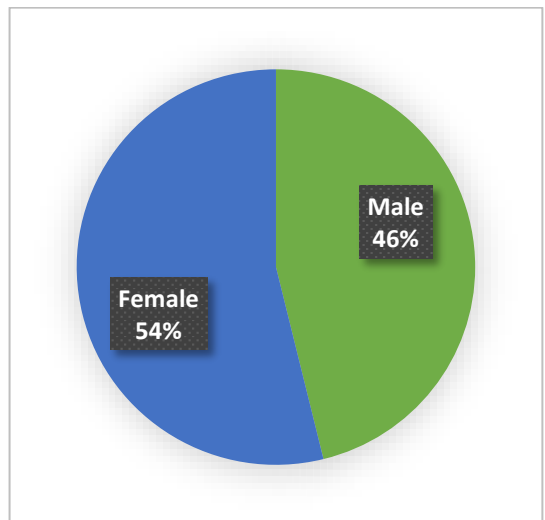

**Fig. 5. Gender wise hyponatremia prediction by MLP**

The prediction results of the proposed MLP algorithm suggests that it predicts the future sodium level of the patients based on their disease history. Therefore, the anticipated benefits of the proposed MLP algorithm will be useful to patients with the similar disease history. Such that the physicians can make use of the proposed MLP model in order to make the appropriate decision on treatments (based on patient disease history and proposed MLP results) and possible outcome of the patients. Based on the patients history and clinical details (gender, age, diabetes, pneumonia, liver_disease, malignancy, pulmonary, sepsis, SIADH) the proposed MLP can able to predict patient's sodium levels.

### 4.4 Computation and analysis of Precision rate

In order to analyse the accuracy of prediction results of the MLR and MLP algorithms, the precision rate is calculated. The prediction precision rate is the percentage of root of squared difference rate between the predicted and observed results calculated using the Euclidean distance [35]. It is given in the equation (2) as,

$$PR = \left\{1 - \left(\frac{EUC_{Dis}}{EUC_{max}}\right)\right\} * 100 \qquad (2)$$

$$EUC_{Dis} = \sqrt{(Predicted - observed)^2}$$

Where, $EUC_{Dis}$ denotes the Euclidean distance and $EUC_{max}$ represents the maximum Euclidean distance among predicted and observed serum sodium results [35]. Table.7 gives the accuracy of the prediction results using precision rate analysis. In Table.7, for the different serum sodium range such as less than 120, 120 to 125, 126 to 130, 131 to 135, and 136 to 138, the proposed MLP algorithms has the prediction precision rates as 98.7928, 90.5073, 95.0537, 98.1052, and 96.9456, whereas the MLP algorithms has the prediction precision rates as 51.2414, 40.5436, 57.1051, 71.6174, and 47.7739. The proposed MLP algorithm has the prediction accuracy of 90.5 to 98.7 %, whereas the MLR algorithm has the prediction accuracy of 40.5 to 71.6 %. Such that the proposed MLP algorithm has 27-50 % of higher precision rate on predicting the future sodium range of the patients.

**Table.7 Precision rate for MLR and MLP**

| Precision Rate (%) | | |
|---|---|---|
| Algorithm/ Na+ Range | MLR | MLP |
| <120 | 51.2414 | 98.7928 |
| 120 to 125 | 40.5436 | 90.5073 |
| 126 to 130 | 57.1051 | 95.0537 |
| 131 to 135 | 71.6174 | 98.1052 |
| 136 to 138 | 47.7739 | 96.9456 |

Table.8 Percentage of Difference for Observed, MLR, and MLP results

| Results/<br>Na+ Range | PD (%) | | |
|---|---|---|---|
| | MLP with<br>observed results | MLR with<br>observed results | MLP with<br>MLR results |
| <120 | 1.2145 | 71.0682 | 72.127 |
| 120 to 125 | 9.0625 | 45.8315 | 37.1548 |
| 126 to 130 | 5.0717 | 54.6066 | 49.8803 |
| 131 to 135 | 1.877 | 24.8553 | 23.0051 |
| 136 to 138 | 3.1018 | 70.6838 | 67.9545 |

Table. 8 gives the summary of percentage of difference (PD) among the observed results, MLR algorithm based prediction results, and MLP algorithm based prediction results. In this analysis the proposed MLP algorithm were produced only 1.2 to 9 % of difference with the observed results; whereas the MLR algorithm based prediction results has 24.8 to 71 % of difference with the observed results. Since, the proposed MLP algorithm were produced 23.6 to 62 % of reduced percentage of difference with the observed results. As well as the improved results by the MLP prediction results as compared with MLR prediction results is 23-72 %.

## 5. Conclusion

This work were concentrated on prediction of future sodium range for the patients based on various health history factors such as age, gender, health problems, etc. in order to predict the hypo/hyper-natremia. The proposed MLP algorithm has produced the accurate future serum sodium prediction range than the MLR algorithm. The MLR algorithm has the prediction accurate rate of 41-72 %, whereas the MLP neural network algorithm has the accurate prediction of 91-99 %. The MLP algorithm based prediction results has 27-50 % of improved prediction accuracy than the MLR algorithm based prediction results. Moreover the proposed MLR algorithm based prediction results 57.1 % of reduced MSE error rate than the MLR results on predicting future sodium ranges of patients. The outcome of the proposed MLP algorithm based future health prediction algorithm could be more helpful for physicians and patients to make further decisions based on their health conditions. Based on these results the future work will concentrate on forecasting the possibility of the occurrence of hyponatremia once again using machine learning prediction algorithms.

**Acknowledgements**

No Funding received